# Dual-Gate GaAs-Nanowire FET for Room Temperature Charge-Qubit Operation: A NEGF Approach

Basudev Nag Chowdhury[1] and Sanatan Chattopadhyay[1,2*]

[1]Department of Electronic Science, University of Calcutta, Kolkata, India

[2]Centre for Research in Nanoscience and Nanotechnology (CRNN), University of Calcutta, Kolkata, India

*E-mail address: scelc@caluniv.ac.in

**Abstract**

The current work investigates the performance of dual-gate GaAs-nanowire FET as a charge-qubit device operating at room temperature. In compatibility with the state-of-the-art classical bit technology, it is shown that the single gate of a nanowire FET can be replaced by two localized gates to achieve such charge-qubit operation. On application of relevant biases to the localized gates, two voltage tunable quantum dots are created within the nanowire channel with electrostatically controlled single-state-occupancy and inter-dot coupling leading to charge-qubit operation at room temperature. The associated electron transport is theoretically modeled on the basis of non-equilibrium Green's function (NEGF) formalism. The 'initialization' and 'manipulation' for qubit operation are performed by applying suitable gate voltages, whereas the 'measurement' is executed by applying a small drain bias to obtain a pulse current of ~pA order. A ~25 MHz frequency of coherent oscillation is observed for the qubit and a characteristic decay time of ~ 70 ns is achieved. The results suggest that such dual gate nanowire FET is a promising architecture for charge-qubit operation at room temperature.

## I. Introduction

The advent of quantum computation and quantum information processing in practice is fundamentally based upon the development of qubits, where quantum superposition principle leads to coherent oscillation between two qubit-states within their characteristic coherence time [1, 2]. Amidst various approaches to develop spin and charge based qubits, the exploitation of SQUID including Josephson junction [3-6] and semiconducting quantum dots (QDs) [7-14] have drawn key attention for achieving desired control over the coherent manipulation of their quantum states. Especially, the double quantum dot (DQD) device with a single excess charge has emerged as one of the most promising qubit architectures due to its electrostatically controlled operation by implementing a number of voltage gates [7, 9-13]. However, a principal challenge for implementing all such device schemes for practical applications is that such qubits can operate only at very low temperature (~ mK). Further, such device schemes for qubits are fundamentally different from the state-of-the-art classical bits generation schemes, which use the matured metal-oxide-semiconductor field-effect-transistor (MOSFET) technology or their emerging evolutionary counterparts, such as, nanowire FETs. Evidently, the relevant mature technology for qubit generation is still a concern in terms of reliability, complexity and cost-effectiveness. Therefore, despite the successful and systematic progress in prototype experiments of the aforesaid approaches, the increasing social drive for quantum computation and quantum information processing demands for the innovative architectures, that needs to be compatible, scalable and integrable with state-of-the-art of matured semiconductor technology.

In this context, the current work proposes a scheme with dual-gate GaAs-nanowire FET architecture for charge-qubit operation at room temperature. The conventional approach of generating classical bits is to use a single gate over the entire channel of a FET. However, it is shown in the current work that charge-qubit operation can be achieved if two localized gates are used over the channel of a nanowire FET instead of a single gate. In such architecture, appropriate biases on the two localized gates can create two voltage tunable quantum dots (VTQDs) with single state each, when the inter-dot tunneling is insignificant. The occupancy of such state of QDs can be controlled by appropriate biases at the respective gates with respect to source/drain. However, the coupling between two QDs can be tuned from the 'non-resonant' to 'resonant' condition by manipulating gate voltages relative to each other, and thereby, either localizing an electron at one of the QDs, or to delocalize it over both the QDs, respectively.

Accordingly, the basic modes of qubit operation such as 'initialization' and 'manipulation' are obtained by varying relevant voltages at the two gates, and the 'measurement' is performed by applying a small bias at the drain terminal. The results of such operations are predicted in the current work by self-consistently solving the relevant quantum-electrostatic equations for such proposed dual-gate nanowire FETs, employing non-equilibrium Green's function (NEGF) approach.

## II. Scheme of device and qubit-operation

The schematic of dual-gate GaAs-nanowire FET device considered in the current work is depicted in Fig. 1(a), where an Ω-gate configuration is assumed, owing to their lesser complexity for fabrication process. The nanowire material is chosen to be zincblende <100>-GaAs due to its symmetric effective mass tensor of low values (m*~0.06), even in the GaAs quantum structures [15, 16], which can lead to create discrete quantized states at room temperature [17]. It is worthy to mention that such GaAs nanowires with polytypism-free zincblende structures have already been grown successfully [18-20], including along <100> direction [18]. In the current nanowire MOSFET, $SiO_2$ of 2 nm thickness is selected as the gate insulator since it has been reported to exhibit minimum gate tunneling leakage current in nanoscale MOS architecture [21]. The source/drain are considered to be GaAs with a doping level of $5\times10^{17}$ /cm$^3$ since such order of doping concentration is sufficient to make it degenerate [22]. In order to achieve strong quantum confinement for realizing a one-level device at room temperature, diameter of the nanowire is taken to be 5 nm and the gate lengths are considered to be 3 nm each, all of which are significantly smaller than the excitonic Bohr radius of GaAs (~12 nm) [17]. At this point, it is imperative to mention that, sub-5 nm gate MOSFETs [23-25], particularly 3 nm technology, have already been developed successfully [26], and is currently in large scale industrial production with FinFET and GAA nanowire FET architectures [27, 28]. In the current work, total channel length of the nanowire FET is considered to be 20 nm that refers to the ballistic transport regime with minimized scattering [29, 30], which is necessary for reducing the associated decoherence effects. Analyzing the penetrability of electron wave from source/drain into the two adjacent VTQDs, the source-to-gate-1 and gate-2-to-drain distances are assumed to be 3 nm and 8 nm, respectively. Consequently, for qubit 'initialization', the two gate voltages can be defined in such a manner that the electron from source can tunnel into VTQD-1 to occupy its single state, however, such single state of VTQD-2 remains unoccupied due to negligible electron tunneling probability through its larger distance from the drain. It is observed that for the present considerations of nanowire dimensions and materials, the electron tunneling probability between reservoirs and their adjacent QDs as well as between the two QDs drops down significantly for ~5 nm separations, which evidently leads to the abovementioned choices of source-to-gate-1 and gate-2-to-drain distances. Limited by such constraints, the distances could be slightly changed (*e.g.*, by ~1 nm), however, for such considerations, only the corresponding gate voltages would be accordingly different from the values in the present case, but not the qualitative nature of the results. Further, such range of variations in separations may be within the error limit in real devices during their fabrication, and therefore of minimal practical significance. Finally, in order to electrically isolate the active device from substrate, the entire device is assumed to be fabricated on an insulator-on-substrate (IOS) platform as schematically shown in Fig. 1(a).

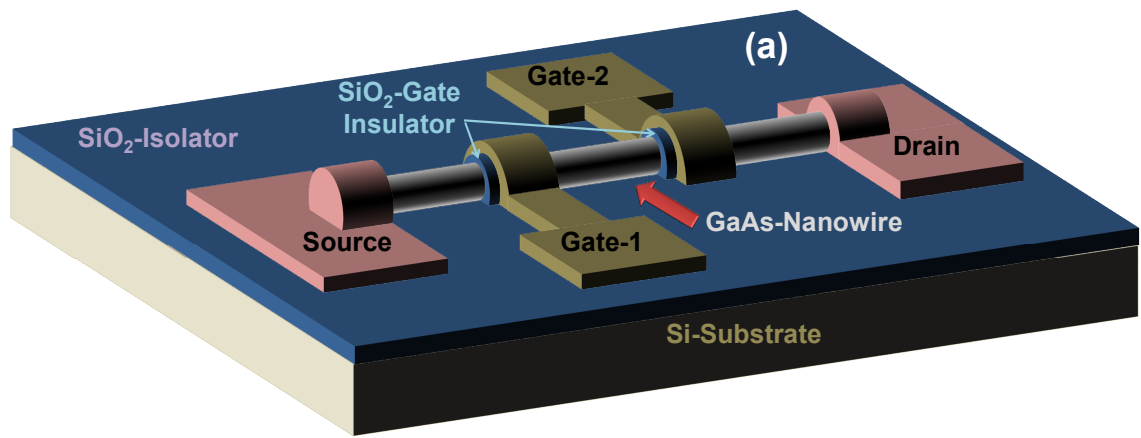

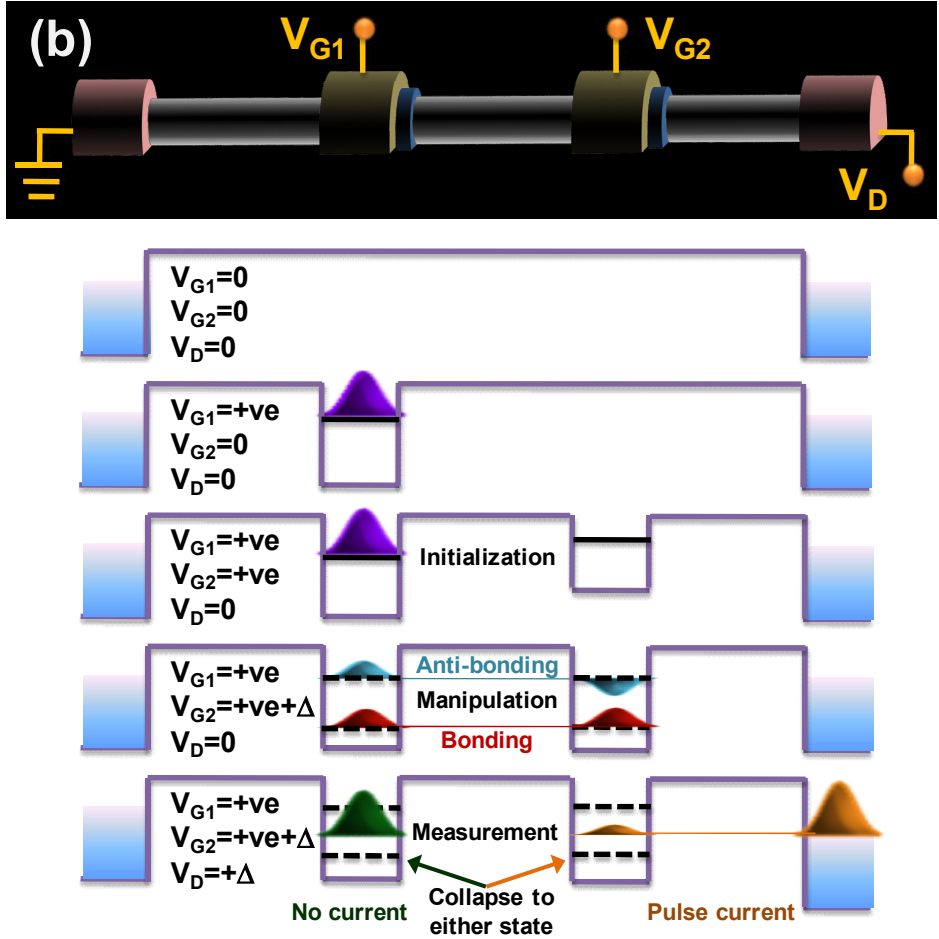

Fig. 1. (a) The schematic of dual-gate GaAs-nanowire FET device on IOS platform; (b) schematic band alignment along the nanowire channel representing the modes of qubit operation by varying the voltages at two gates and drain: first, no bias is applied on the device; second, a voltage is applied to gate-1 to create QD-1 so that a single electron can enter from source to its ground state; third, voltage is given to gate-2 to create QD-2 but no electron enters to it from QD-1 or drain; fourth, to vary the voltage at gate-2 to create and manipulate superposition between two QDs leading to the development of 'bonding' and 'anti-bonding' states; and finally, to apply a small drain bias to collapse the superposition resulting in 'No current' or a 'Pulse current' at the drain.

Further, the phase breaking effect in the device current due to phonon scattering, which might be the another source of decoherence, has already been shown to be negligibly small in a one-level system [31] and therefore it is not considered in the present work. It is also imperative to mention that the energy gap between ground state and first excited state in the present VTQDs with considered materials and dimensions is obtained to be ~500 meV (for detail, see **Appendix-I**), and thus, thermal voltage at room temperature (~25 mV) is insufficient for corresponding excitation associated with the phonon induced intra-band scattering (for detail, see **Appendix-II**). Moreover, charge-qubit operation in the present device architecture is associated with the lowest unoccupied state (*i.e.* single ground state only) unlike the conventional DQD charge-qubits that exploit highest unoccupied state with already occupied lower sates. Therefore, conventional noise in charge-qubits is expected to be minimized in the present dual-gate nanowire FET device structure.

Fig. 1(b) represents the scheme of qubit operation in such a dual-gate nanowire FET device. When appropriate voltages are applied to the gates (*i.e.*, $V_{G1}$ at gate-1 and $V_{G2}$ at gate-2), two 3D-quantized potential wells with single state each are created locally within the channel underneath the gates, which are referred to $|L\rangle$ and $|R\rangle$, respectively. Due to positional asymmetry of the gates (*i.e.*, difference in source-to-gate-1 and gate-2-to-drain distances), $|L\rangle$-state is initially occupied due to the inflow of an electron from source whereas $|R\rangle$-state remains unoccupied. The gate biases can then be tuned to manipulate the inter-dot tunnel coupling to get resonance by creating a superposed state, $|\psi\rangle = \alpha(V_{G1},V_{G2})|L\rangle + \beta(V_{G1},V_{G2})|R\rangle$, where $|\alpha|^2$ and $|\beta|^2$ are probabilities of the electron to be within the corresponding QDs. If a small drain bias ($V_D$) is applied at such condition, the electron will solely be in either of the QDs suffering collapse to $|L\rangle$-state or $|R\rangle$-state, resulting to 'no drain current' or 'a pulse current' (~pA order for the present consideration of device dimensions and materials [10]), respectively, leading to qubit oscillation. Such order of current at room temperature can be measured by a suitable semiconductor characterization setup [32]. It may be worthy to note that the collapse of such superposed state can also be executed by measuring one of the gate capacitances, which is of the order of ~aF. Such a small capacitance can be measured by following the approach reported in [33]. Further, the degeneracy of the superposed state splitting into 'bonding' and 'anti-boding' states is discussed later in the present work.

### III. Theoretical modeling

The general transport properties of nanowire FET devices have been extensively studied by developing the relevant NEGF formalism [34-38]. Following such approach, the qubit operation using the present dual gate nanowire FET is theoretically modeled by considering Hamiltonian of the nanowire channel coupled with source and drain as,

$$H = \sum_i H_{Iso} c_i^+ c_i + \sum_{i,r} \left( \tau_{ir}^{S/D} c_i^+ c_r + \tau_{ri}^{S/D} {}^* c_r^+ c_i \right) \quad (1)$$

where, $H_{Iso}$ is the Hamiltonian of isolated nanowire when it is not coupled to the source/drain (for detail, see **Appendix-I**), whereas the couplings between $i^{th}$ state of nanowire channel and $r^{th}$ state of source/drain are given by $\tau_{ir}^{S/D}$. The second quantization field operators $c_i^+$ and $c_i$ (or $c_r^+$ and $c_r$) represent creation and annihilation of electrons within the nanowire channel (or in source/drain), which follow the Fermi-Dirac anti-commutation relations,

$$\begin{cases} \{c_i, c_j^+\} = \delta_{ij} \\ \{c_i, c_j\} = 0 \\ \{c_i^+, c_j^+\} = 0 \end{cases} \quad (2)$$

indicating single occupancy of each quantum state for electrons obeying the exclusion principle. It is apparent that the source/drain consists of a large a number of electrons at several energy states, whereas the gate voltage-assisted quantum dots within the nanowire channel have single state each with single occupancy. Therefore, the low-energy field theory including fermionic second quantization is necessary to model such one-to-many state coupling along with single occupancy. However, Kramers degeneracy to consider the probability of two electrons (with opposite spins) to occupy such a single state is incorporated into the model by considering the relevant expressions of local density of states (LDOS) and device current as follows.

The relevant equations of motion for electrons are derived from the Hamiltonian using Heisenberg equation and can be solved in 'coupled mode space' (for detail, see **Appendix-I**) from the local non-equilibrium Green's function as a function of electron energy ($E$) in the nanowire coupled to source/drain as given by [34-36],

$$G(E) = [E - H_{Iso} - \Sigma_S(E) - \Sigma_D(E)]^{-1} \quad (3)$$

where, $\Sigma_S$ and $\Sigma_D$ are the self-energies of source and drain, respectively, which depend on the corresponding coupling strength, $\tau_{ir}^{S/D}$ [36]. The LDOS occupied by electrons are therefore obtained to be,

$$D(E) = \frac{i}{2\pi a}\left[\left(G(E)[\Sigma_S(E) - \Sigma_S^+(E)]G^+(E)\right)f_S(E) + \left(G(E)[\Sigma_D(E) - \Sigma_D^+(E)]G^+(E)\right)f_D(E)\right] \quad (4)$$

where $f_S$ and $f_D$ are Fermi-Dirac distribution functions corresponding to source and drain, and '$a$' is the grid spacing (i.e., half the lattice constant as considered for present simulation). It is apparent from Eq. 4 that when the coupling of VTQDs with source/drain be negligibly small (i.e., $\Sigma_{S/D}(E) \to 0$), the density of states tends to twice the delta function. Therefore, the expression for occupied LDOS of Eq. 4 includes Kramers degeneracy to inherently consider the probability of two electrons (for each spin) to occupy a single state. However, despite such availability of energy space for two electrons at a single state of the QDs, whether it will be occupied by two electrons, or by one only, or will be left unoccupied, depends on the position of QD-eigenstate in energy space relative to the Fermi levels of source/drain. It is further worthy to mention that although the spin degeneracy is included in the model, the spin-orbit coupling is not considered in the current work, since its effects in GaAs quantum structures are reported to be observed only at very low temperatures [39-41].

At this point it is imperative to mention that, integrating the occupied LDOS over energy, followed by its normalization in coordinate space (suppose $z$-axis along nanowire channel), gives rise to the positional probability ($|\psi(z)|^2$) of electron within the device. For numerical analysis, the carrier profile ($n(z)$), obtained by assuming an initial potential distribution in isolated Hamiltonian ($H_{Iso}$), is put into the Poisson's equation to get resultant potential distribution along the channel ($\varphi(z)$) including two VTQDs in an iterative method to get self-consistency is achieved. It is worthy to mention that Poisson's equation for the nanowire FET with an $\Omega$-gate structure leads to [38],

$$\left[\frac{d^2}{dz^2} - \frac{2}{R^2}\frac{\varepsilon_{Ox}}{\varepsilon_{NW}}\frac{1}{\ln(1 + t_{Ox}/R)}\right](\varphi(z) - V_G(Z)) = \frac{e}{\varepsilon_{NW}(\pi - \theta)R^2}n(z) \quad (5)$$

where, $\varepsilon_{NW}$ and $\varepsilon_{Ox}$ are permittivity of the nanowire and gate oxide, respectively; $t_{Ox}$ is oxide thickness and $R$ is the nanowire radius; $V_G(Z) = V_{G1}$ and $V_G(Z) = V_{G2}$ throughout the gate-1 and gate-2, respectively; and $\theta = \cos^{-1}(R/(R+t_{Ox}))$ for Ω-gate without any insertion of nanowire into the IOS [38]. Finally, on application of a small drain bias ($V_D$), the amplitude of the pulse current is obtained to be [36],

$$I_0 = \frac{2e}{h}\int Trace\big(i[\Sigma_S(E) - \Sigma_S^+(E)]G(E)i[\Sigma_D(E) - \Sigma_D^+(E)]G^+(E)\big)\big(f_S(E) - f_D(E - V_D)\big)dE$$

(6)

where, the factor '2' multiplied with quantum conductance represents Kramers degeneracy. The oscillatory decay of such pulse in time, originated due to coherent oscillation of qubit, can be obtained from the retarded Green's function following [37] as (for detail, see **Appendix-III**),

$$I = I_0\big(F.T._{E \to t}[G_{Iso}(E)\Sigma_D(E)G(E)]\big) \qquad (7)$$

where $(F.T._{E \to t})$ indicates Fourier transform from energy space to the time domain and $G_{Iso}(E)$ represents the Green's function for isolated channel when its coupling with source/drain is negligibly small.

## IV. Results and discussion

The performance of dual gate GaAs nanowire FET considered in the present work as a charge-qubit is studied from the results obtained by numerically solving the above equations in Matlab. In such method of computation, first the Hamiltonian for active device (*i.e.*, the nanowire channel) is constructed assuming an arbitrary potential distribution and the device Green's function is obtained by incorporating self-energies for the reservoirs (source/drain). The local density of states (LDOS) at each point is calculated from such Green's function, yielding the carrier distribution, which is then put into the Poisson's equation to compute potential profile. Such potential values at each point are then put again into the device Hamiltonian and this process is run further. The iteration is continued until the input and output potential values become identical within the limit of accuracy, when self-consistency is considered to be achieved [38]. The resultant conduction band profile indicating the formation of voltage tunable double quantum dots in the present dual gate nanowire FET is shown by dotted lines in Fig. 2. It is worthy to mention that, physically, both the potential and charge distributions are outcome of the complementary effects of quantum (given by Schrodinger equation) and electrostatic (given by Poisson equation) properties of the electrons. As a simultaneous result of these dual effects, a sharp potential barrier and strong quantization is obtained in the present device, beneath each gate at a corresponding high gate voltage (~ 1 V along 5 nm diameter). However, it is also imperative to note that at such voltages, due to the strong confinement, the ground state is at much higher position (*i.e.* >500 meV, as can be seen in Fig. 2) from the conduction band. Therefore, even such voltages are not expected to result in electronic transitions to conduction band of QDs from the valence band of the adjacent regions.

Further, Fig. 2 depicts its basic modes of operation including 'initialization', 'manipulation' and 'measurement'. The variation of occupied LDOS in energy space along the nanowire channel consisting of two VTQDs at such operational modes is plotted in Fig. 2. Prior to the discussion on Fig. 2, it is worthy to note that, the Hamiltonian and corresponding wavefunction for the coupled VTQDs are not solved in the current work in the $(|L\rangle, |R\rangle)$-discrete 'basis' states, but in 'coupled mode space' using the 'basis' of a continuous position state along the nanowire axis and discrete energy sub-band states for transverse directions (for detail, see **Appendix-I**). However, the wavefunctions extended up to the boundary of either of the 'Left' and 'Right' QDs can be obtained from the LDOS, as discussed earlier. When the inter-dot coupling is negligibly small, such wavefunctions represent the positional projection (along nanowire axis) of the $|L\rangle$ and $|R\rangle$-states for each transverse sub-band mode. On the other hand, finite inter-dot coupling leads to superposition of such sub-band states of the two QDs, which can be manipulated by applying appropriate biases at the two gates and are described by unitary evolution. However, the application of a small bias at the drain probabilistically results in a non-equilibrium ballistic outflow of the electron resulting in the non-unitary collapse of such superposition. Therefore, such process cannot be solved by considering the Hamiltonian only, and is modeled in the current work by considering the corresponding self-energy ($\Sigma_D(E)$).

In Fig. 2, the dotted lines represent calculated values (from the present NEGF model) of the self consistent conduction band profile along nanowire channel for the corresponding biases. The 'initialization' ($|\psi\rangle = |L\rangle$) condition is obtained by applying a voltage of 1.15 V at gate-1 and 1.3 V at gate-2 without any bias at the drain. It is apparent from the plot of Fig. 2 that asymmetric position of the gates enables to create a significant difference between the geometry of voltage-assisted 3D-quantized potential wells of the two QDs and thereby makes it possible for electron to occupy the single state of QD-1 keeping the state of QD-2 empty which is essential for such 'initialization'.

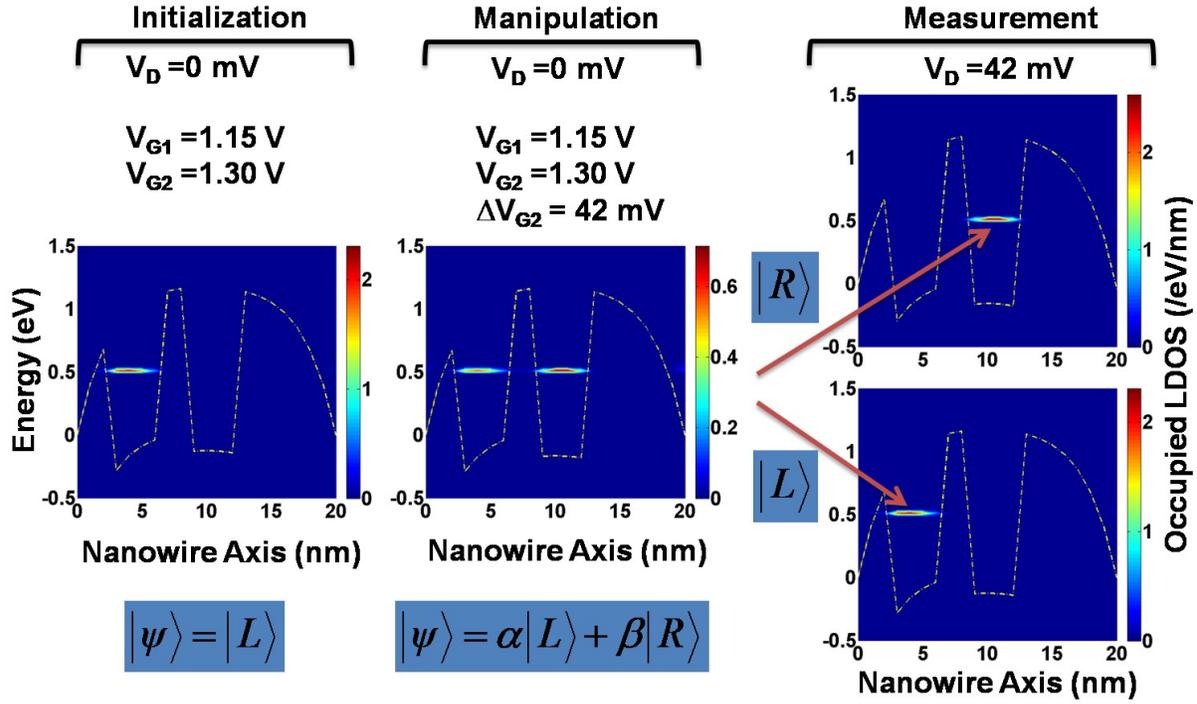

Fig. 2. Plots of the variation of occupied LDOS (contour) in energy space along the nanowire channel with conduction band alignment (dotted line) showing the two VTQDs at different modes of qubit operation.

If an incremental pulse voltage ($\Delta V_{G2} \sim$ mV) is now applied on gate-2, the superposed state $|\psi\rangle = \alpha|L\rangle + \beta|R\rangle$ is obtained, where the single electron is partially at QD-1 and partially at QD-2 at the same time due to inter-dot resonance tunneling. The corresponding probabilities, $|\alpha|^2$ and $|\beta|^2$, can be manipulated by varying the pulse voltage that modulates the broadening of resonance; and almost an equal probability in both QDs is obtained at $\Delta V_{G2} = 42$ mV shown in Fig. 2. At such condition, a small drain bias of 42 mV is sufficient for the collapse of superposed state into either of $|R\rangle$ or $|L\rangle$ and accordingly it creates either a 'pulse current' or 'no current' through the device. It is imperative to note that the drain voltage must be small enough so that

the electric field along the nanowire axis does not influence the result of measurement toward collapsing into the state $|R\rangle$ in comparison to $|L\rangle$.

The manipulation of superposition of the states $|L\rangle$ and $|R\rangle$ is demonstrated in Fig. 3(a), which represents the positional probability of electron, delocalized over the two VTQDs, for the pulse voltage at gate-2 varying in the range of 30-48 mV. It is apparent from such plots that probability of $|L\rangle$-state decreases with increasing $\Delta V_{G2}$ in the considered range, leading to a corresponding increase of the probability of $|R\rangle$-state. A close inspection of the results shown in Fig. 3(a) indicates the positional probability distribution in QD-1 to be asymmetric with the maxima being nearer to the source, which is attributed to the asymmetric well shape of QD-1 due to conduction band bending near the source (Fig. 2). Such position of the maxima may be beneficial for reducing the possible source induced decoherence by Coulomb blockade effect **[7, 9]**.

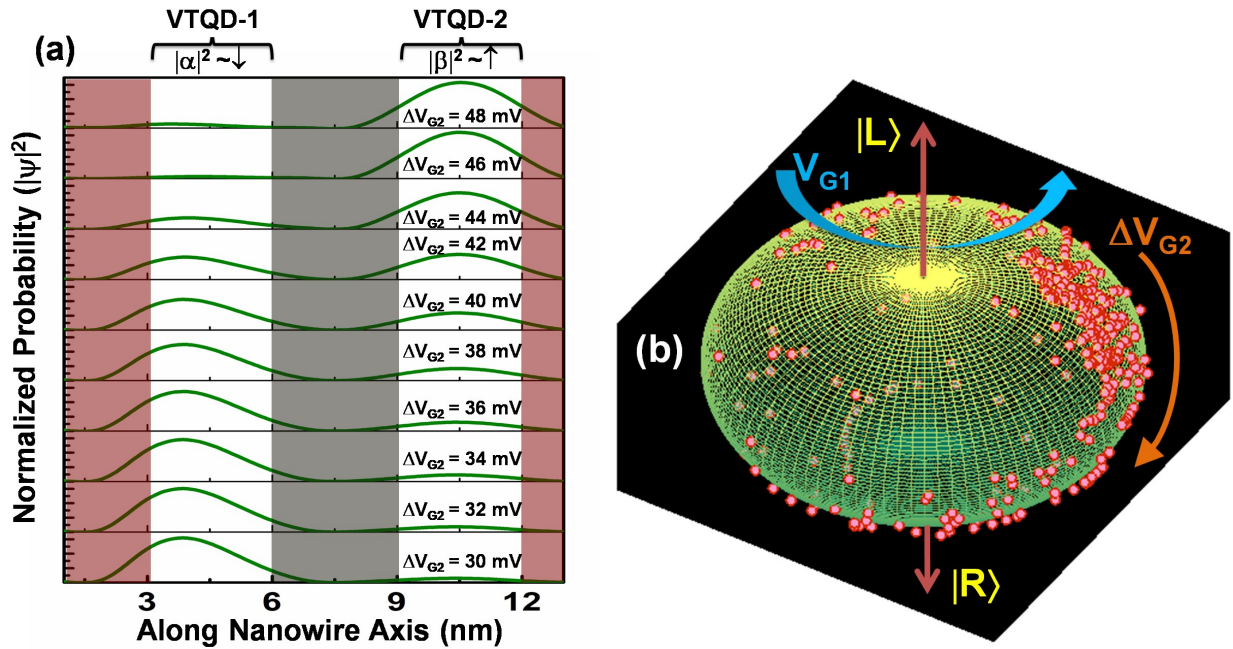

Fig. 3. (a) The plots of positional probability of electron, delocalized over the two VTQDs, for the pulse voltage at gate-2 leading to manipulation of superposition of the states; (b) Bloch sphere representing the superposed states $|\psi\rangle = \alpha|L\rangle + \beta|R\rangle$ varying $V_{G1}$ (1.12-1.16 V) and $V_{G2}$ (1.33-1.348 V) at $V_D$=0 V; the dotted colour points represent the phases during qubit manipulation.

The manipulation of qubit in terms of relative amplitude along with phase of the two states is represented by plotting the corresponding Bloch sphere, as shown in Fig. 3(b), where the north and south poles indicate $|L\rangle$ and $|R\rangle$ states, respectively. Since $|\alpha|$ and $|\beta|$ can be manipulated by changing the bias at gate-2 for a given voltage at gate-1 under the normalization constraint as shown in Fig. 3(a), the polar angles are obtained by varying the $\Delta V_{G2}$ in the range of 30-48 mV around the operating point (*i.e.*, $V_{G1}$ = 1.15 V, $V_{G2}$ = 1.342 V). On the other hand, the different azimuthal angles represent relative phases between the states $|L\rangle$ and $|R\rangle$ which are resulted from small variation of $V_{G1}$ in the range of 1.12-1.16 V. Such azimuthal angles are calculated from the phase difference of local values of Green's function at the two QDs, where the phase values at each point of the QDs are obtained by integrating the occupied local Green's function over the entire energy range. The differences in such phase values between the analogous points (*e.g.*, the starting/central/ending grid) of the two QDs are then calculated. Such inter-dot phase differences at different positional coordinates within the QDs (*e.g.*, the starting/central/ending grid) are observed to be very small, which is attributed to the small and almost identical dimensions of the two QDs.

At this point, it is worthy to mention that both the gate voltages are associated with 'initialization' and 'manipulation' process in the proposed qubit depending on their absolute and relative values. For instance, the considered geometry, dimension and materials of the current device leads to occupancy of the ground state of QD-1 at a minimum $V_{G1}$ ~ 1.1 V, whereas it requires $V_{G1}$ ~ 1.6 V to occupy its first excited state. In such condition, the ground state of QD-2 remains completely unoccupied up to $V_{G2}$ < 1.3 V; and the electron, occupying the ground state of QD-1, delocalizes up to QD-2 for $V_{G2} \geq 1.3$ V. Such delocalization can be manipulated by a relevant increment of $\Delta V_{G2}$ (*i.e.* 30-48 mV here) which increases the fractional probability of the electron in QD-2 with respect to QD-1 (as shown in Fig. 3(a)), thereby rotating the Bloch vector along polar angle from North Pole ($|L\rangle$) toward South Pole ($|R\rangle$). On the other hand, the azimuthal angle, which represents the phase of electron in QD-2 with respect to QD-1, varies with $V_{G1}$ (*i.e.* here 1.12-1.16 V) indicating the rotation of Bloch vector around Pole-axis, until the $V_{G1}$ is high enough (*i.e.* here ~1.6 V) to result a further inflow of electron from source to occupy the first excited state of QD-1.

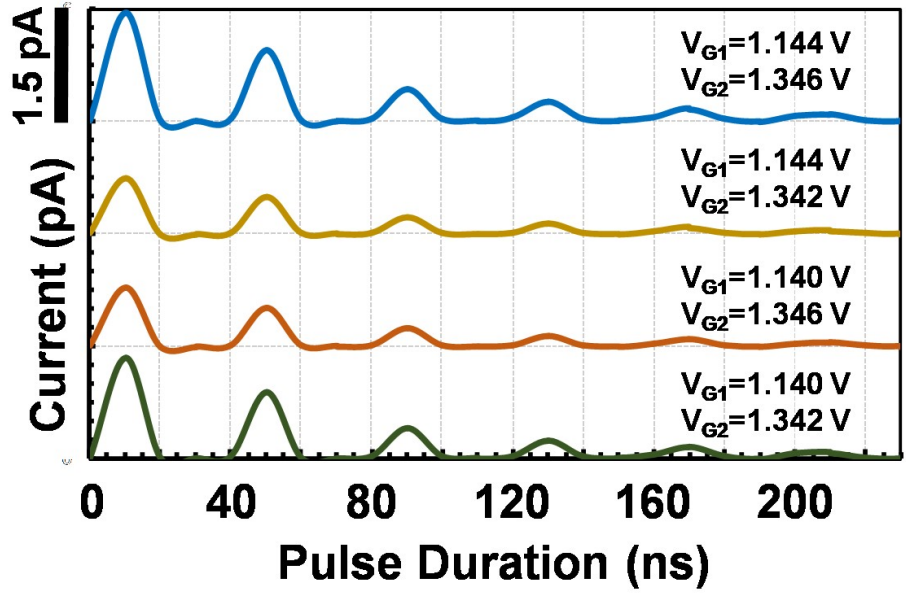

Fig. 4. The variation of output signal (drain current) with time at different operating points defined by the corresponding $V_{G1}$-$V_{G2}$ combinations with $V_D$=42mV, indicating coherent oscillation enveloped by qubit dephasing.

The manipulation of such superposed state, in time domain, can be conducted by applying $\Delta V_{G2}$ as the pulses with its duration less than the overall pulse repetition time [11]. Such repetition time in the present dual gate nanowire FET device is considered to be the average lifetime of the electron pulse, entering QD-1 (for detail, see **Appendix-III**), and obtained to be ~336.5 ns. Therefore, the pulse duration of $\Delta V_{G2}$ is varied up to ~250 ns and the corresponding response of pulse current is studied with a combination of voltages at gate-1 and gate-2 around the operating point, and plotted in Fig. 4. The qubit oscillation frequency is found to be ~25 MHz, exhibiting a negligible variation for the considered values of $V_{G1}$ and $V_{G2}$. Such coherent oscillation frequency of the qubit generated by present dual gate nanowire FET is of the order of previously reported results [11]. The characteristic decay time for qubit dephasing is obtained to be ~73 ns (for detail, see **Appendix-III**), which is almost 1/3$^{rd}$ of the reported value for Si-DQD at a temperature of ~20 mK, however, significantly higher than the results (~1 ns and less) obtained for GaAs/AlGaAs based DQD structure at ~100 mK [9, 10].

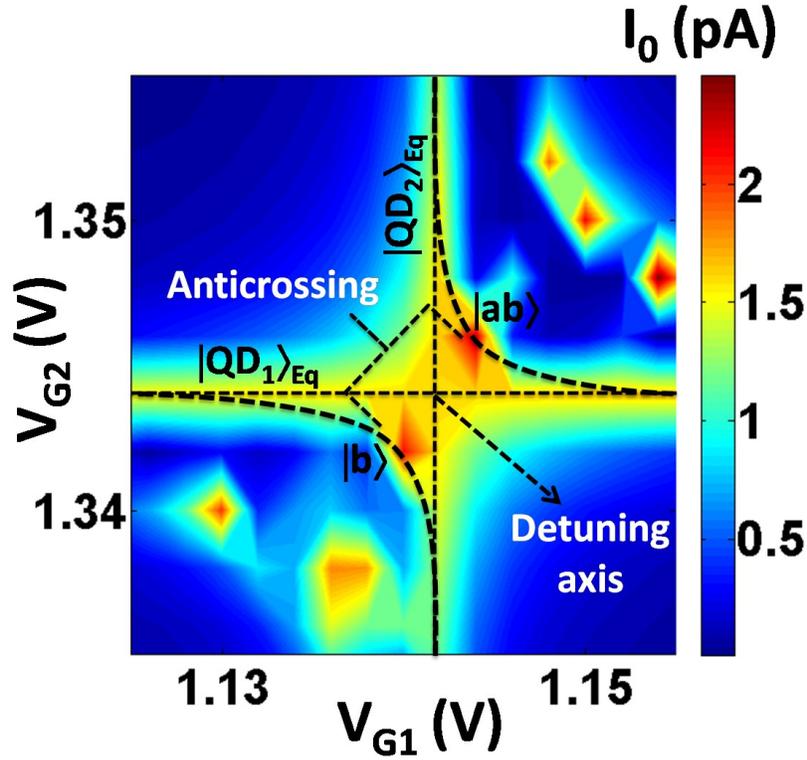

Fig. 5. Mapping of output signal amplitude (maximum drain current) on the $V_{G1}V_{G2}$-plane at $V_D$=42 mV. The dotted hyperbolic curves represent the 'bonding' ($|b\rangle$) and 'anti-bonding' ($|ab\rangle$) states during resonance and dotted lines indicate such states of equivalent uncoupled QDs ($|QD_1\rangle_{Eq}$ and $|QD_2\rangle_{Eq}$).

It is imperative to mention that the operating point for maximum amplitude of the output signal ($I_0$) is observed to shift marginally in both the gate voltage axes due to the applied drain bias for 'measurement', and it splits into two states, *i.e.* ($V_{G1},V_{G2}$) =(1.14 V, 1.342 V) and (1.144 V, 1.346 V). This is attributed to the fact that during inter-dot resonance tunneling while manipulating the qubit superposition, the single electronic state splits into two states, namely, 'bonding' and 'anti-bonding', distributed over both the QDs [7]. The energy variation of such 'bonding' and 'anti-bonding' states is a function of energy difference between the isolated states (*i.e.* detuning) of the two QDs and their anticrossing energy for coupling [7, 11]. Such effect is manifested in the mapping of output signal amplitude on the plane of two gate voltage axes, known to be the stability diagram [42], as shown in Fig. 5. Mappings of similar nature in the stability diagram are reported in a number of experimental results obtained in DQD structures **[10, 43-45]**.

Fig. 5 indicates that, the 'bonding' and 'anti-bonding' states of coupled-QDs ($|b\rangle$ and $|ab\rangle$) and equivalent uncoupled QDs ($|QD_1\rangle_{Eq}$ and $|QD_2\rangle_{Eq}$) as a function of their detuning and the anticrossing energy **[11]** are expressed as the output current variation with two gate voltages through a coordinate transformation (*i.e.* a clockwise rotation). This is schematically traced in Fig. 5 by the dotted hyperbolic curves and straight lines, respectively. In fact, the broadened hyperbolic curves with two gaped vertices represent resonant tunneling current separating the regions where isolated QD-states are not aligned (*i.e.* exhibiting no tunneling current). Such gaped vertices are originated in the stability diagram due to moderate inter-dot coupling **[42]**, the gap being measured by the anticrossing energy **[43, 44]**. However, negligibly weak coupling leads to convert the hyperbolic lines into corresponding straight lines intersecting at an un-split vertex **[42]**. Therefore, the stability diagram represents an equivalent band-structure of the double-QD system, the two gaped vertices indicating band maxima and minima of the bonding and anti-bonding states, respectively. It is worthy to note that the resonant current levels in stability diagram exhibits broadening due to inter-dot coupling as well as the coupling with source/drain. Further, Fig. 5 also shows the appearance of 'excited' 'bonding' and 'anti-bonding' states (below $|b\rangle$ and above $|ab\rangle$) in the present device, which are attributed to the asymmetry in quantum well-shape of the two VTQDs.

## V. Conclusion

To summarize, the present work shows that a replacement of the single gate of a nanowire FET by two localized gates can create charge qubit that is compatible with the state-of-the-art classical bit generation technology. The selection of GaAs as nanowire material and its appropriate dimensions can enable such charge-qubit operation involving single electron state at room temperature. The qubit operation is theoretically modeled by developing a quantum-electrostatic simultaneous solver based on non-equilibrium Green's function (NEGF) formalism. The operational scheme and the obtained values of coherent oscillation frequency and characteristic decay time of the proposed qubit suggest that such dual gate nanowire FET architecture is promising for room temperature charge-qubit operation.

**Acknowledgement**

The authors would like to thank the Center for Research in Nanoscience and Nanotechnology (CRNN), University of Calcutta, for providing the necessary infrastructure support. The authors also acknowledge infrastructural support obtained from the WBDITE, West Bengal, sponsored project.

**Appendix-I: 'Basis' in 'coupled mode space'**

The equations of NEGF formalism for modeling the carrier transport through a nanowire can be solved in 'coupled mode space' instead of 'real space' for numerical simulation to provide highly accurate solutions with considerably higher computational efficiency [46, 47]. In such method, quantum physics is applied to increase the computational efficiency for simulating a quantum structure using classical computational logics. For instance, in a nanowire FET, the electrons are confined in the transverse directions resulting in respective discrete energy sub-bands (*i.e.*, modes); however, they are free to move along the nanowire axis. Thus physically, the description of any cross-section of such nanowire in the 'continuum' of transverse positional coordinates can be replaced by its 'discrete' transverse energy modes. Since only the modes near and below source/drain Fermi level are occupied, in simulation such few numbers of discrete energy modes can replace much higher numbers of grids to represent the continuous positional coordinates, which, in turn, leads to increase the computational efficiency significantly, particularly within the ballistic limit where carrier scattering phenomena are less probable.

Within ballistic regime, the Hamiltonian of an isolated nanowire, *i.e.*, when not coupled to the source/drain, is given under effective mass approximation by [36, 47],

$$H_{Iso} = \frac{1}{2}\sum_{j,k} \hat{p}_j \left(m^{*-1}\right)^{jk} \hat{p}_k + (-e)\varphi(x,y,z) \tag{A1-1}$$

The nanowire is assumed to be composed of several grids along its axis (*i.e.*, $z$-coordinate), and within a particular grid (*i.e.*, where $z$ does not vary), the local transverse Hamiltonian becomes,

$$H_T\big|_z = \frac{1}{2}\sum_{j,k=1}^{2} \hat{p}_j \left(m^{*-1}\right)^{jk} \hat{p}_k + (-e)\varphi(x,y)\big|_z \tag{A1-2}$$

leading to the local transverse sub-band states at each axial point given by,

$$H_T|_z|n_T\rangle_z = E^{Sub}_{n_T}(z)|n_T\rangle_z \tag{A1-3}$$

The state ($|\psi\rangle$) of electron in the 3-D nanowire can thus be expressed in the 'basis' of local transverse sub-band states (for detail see [47]) as,

$$|\psi\rangle = \sum_{n_T} C_{n_T}(z)|n_T\rangle_z \tag{A1-4}$$

which, since Eq. (A1-1)-(A1-3) and using the orthonormality property of transverse sub-band states, gives rise to the equation for isolated nanowire [47],

$$EC_{n_T}(z) = \sum_{\bar{n}_T}\left[H_{n_T,\bar{n}_T}(z) + E^{Sub}_{n_T}(z)\delta_{n_T,\bar{n}_T}\right]C_{\bar{n}_T}(z) \tag{A1-5}$$

When the nanowire is coupled with source/drain, Eq. (A1-5) is modified to Eq. (1) and solved by the Green's function as Eq. (3). It is interesting to note that if the nanowire is grown along one of the principal axes of the material (*i.e.*, <100> zincblende GaAs, as considered in the current work), Eq. (A1-1) and Eq. (A1-2) become diagonalized in terms of the symmetric effective mass tensor. The quantum states in potential well (estimated from the LDOS as given by Eq. (4)), created at a given gate voltage, are obtained by considering the values of electron effective mass components in GaAs, particular diameter of the nanowire and relevant localized gate length. In the present device architecture considered, where nanowire diameter is 5 nm and gate length is 3 nm, the energy spacing between ground state and 1$^{st}$ excited state is obtained using such model to be ~500 meV. However, it can be roughly estimated by considering a rectangular potential well of almost same dimensions (*i.e.*, 5×5×3 nm$^3$), where using the conventional formula $\varepsilon_{n_1,n_2,n_3} = \frac{\hbar^2\pi^2}{2m^*m_0}\left(\frac{n_1^2}{L_1^2} + \frac{n_2^2}{L_2^2} + \frac{n_3^2}{L_3^2}\right)$, the energy spacing between ground state and 1$^{st}$ excited state is $(\varepsilon_{2,1,1} - \varepsilon_{1,1,1}) = 0.7525 \times \left[\frac{0.06}{m^*}\right]$ eV (since $\frac{\hbar^2\pi^2}{m_0}\left(\frac{1}{nm^2}\right) = 0.7525$ eV). For an effective mass of $m^* = 0.06$ (which is the electron effective mass in <100> zincblende GaAs), such spacing becomes ~750 meV. However, exact calculation using NEGF formalism used in the current work results in the value of such spacing to be a little lower than this, *i.e.*, ~500 meV. Therefore, if the materials are selected appropriately, then the VTQDs may have larger energy spacing than that conventionally obtained, and consequently, for such larger energy spacing, the confinement effects can be observed at room temperature **[33, 48]**.

**Appendix-II: Impact of phonon scattering**

In the presence of electron-phonon scattering, Eq. (1) is modified by **[37, 38]**,

$$H = \sum_i H_{Iso} c_i^+ c_i + \sum_{i,r} \left( \tau_{ir}^{S/D} c_i^+ c_r + \tau_{ri}^{S/D} * c_r^+ c_i \right) + \sum_{i,j,\alpha} \left( \tilde{\tau}_{ij}^{\alpha} c_i^+ c_j b_\alpha + \tilde{\tau}_{ji}^{\alpha} * c_i^+ c_j b_\alpha^+ \right) \quad \text{(A2-1)}$$

where $b_\alpha$ is the bosonic field operator for phonons at $\alpha$-mode with angular frequency of $\omega_\alpha$ given by,

$$i\hbar \frac{d}{dt} b_\alpha = \hbar \omega_\alpha b_\alpha + \sum_{i,j} \left( \tilde{\tau}_{ij}^{\alpha} * c_i c_j^+ + \tilde{\tau}_{ji}^{\alpha} * c_j c_i^+ \right) \quad \text{(A2-2)}$$

The phonon interaction potential is $\tilde{\tau} = D\Xi$; $D$ being the deformation potential of the nanowire material and $\Xi = \vec{\nabla} \cdot \vec{u}$ being the lattice strain due to phononic vibrations, given by the lattice displacement of the oscillating atoms with wave vector $\vec{\beta}(\omega)$ as,

$$\vec{u} = \vec{u}_0 e^{i(\vec{\beta} \cdot \vec{r} - \omega t)} + \vec{u}_0 * e^{-i(\vec{\beta} \cdot \vec{r} - \omega t)} \quad \text{(A2-3)}$$

where $\vec{u}_0$ can be approximated considering an Einstein-Debye oscillator as,

$$\vec{u}_0 = \hat{v} \sqrt{\frac{\hbar \omega f_{BE}(\omega, T)}{2\rho \Omega \omega}} \begin{pmatrix} \frac{1}{\sqrt{2}} \\ \pm \frac{1}{\sqrt{2}} \end{pmatrix} \quad \text{(A2-4)}$$

where, $\hat{v}$, $\rho$ and $\Omega$ are the polarization vector, mass density and normalization volume, and $f_{BE}(\omega, T)$ is the Bose-Einstein distribution function for phonons at a temperature $T$. The '+' and '-' signs represent acoustic and optical phonons, respectively. Due to such electron-phonon interaction, the Green's function of Eq. (3) is modified by,

$$G(E) = [E - H_{Iso} - \Sigma_S(E) - \Sigma_D(E) - \Sigma_{ph}(E)]^{-1} \quad \text{(A2-5)}$$

and accordingly, the occupied LDOS for phonon scattering within the nanowire channel consisting of two VTQDs is given by,

$$D_{Sc}(E) = \frac{i}{2\pi a} G(E)\left(\Sigma_{ph} - \Sigma_{ph}^{+}\right) G^{+}(E) \qquad (A2\text{-}6)$$

The model can be used to incorporate the electron-phonon scattering phenomena in the device considered.

At this point, it is worthy to mention that electron-phonon scattering phenomena in quantum devices coupled to reservoirs can lead to atypical electrical characteristics, which are completely unusual compared to the conventional dissipating effects observed at classical/semi-classical devices. For instance, due to sub-band misalignment with reservoirs, phonon scattering can enhance the current in a quantum device instead of conventional degradation in bulk devices **[37]**. Particularly, the effects of phonon scattering in quantum structures can be reduced, even up to a negligibly small level, by increasing the energy gap between sub-bands beyond the phonon energy.

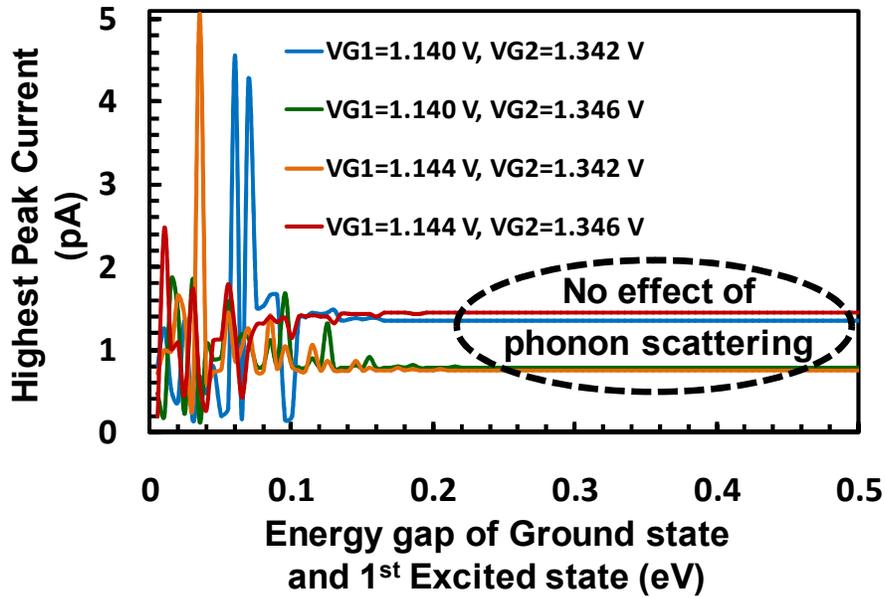

Fig. A2.1. The plots of current in the proposed device structure at room temperature considering electron-phonon scattering with energy gap between ground state and the 1$^{st}$ excited state varying from a (virtually assumed) sub-thermal energy (*i.e.* < 25 meV) to the actual energy gap (*i.e.* ~500 meV).

Such reduction of phonon effects depending on sub-band energy gaps at room temperature may occur at quantum scale, since the phase breaking phenomena associated with 'intra-band' transitions in the 'energy continuum' of bulk materials are transformed into 'inter-sub-band' transitions in 'discrete energy' spectrum of quantum devices. For a comprehensive understanding, the variations of room temperature current at the operating voltages in present device structure due to electron-phonon scattering are plotted in Fig. A2.1. The energy gap between ground state and the 1$^{st}$ excited state is assumed to virtually vary from a sub-thermal energy value (*i.e.*, < 25 meV) to the actual energy gap (*i.e.*, ~500 meV) in the device as obtained in the present work.

The present calculations are performed on the basis of theoretical model discussed above considering both acoustic and optical phonons, where the force constant and deformation potential values for GaAs are taken from [49] and [50], respectively. It is imperative to mention further that such model to study the effect of electron-phonon scattering in semiconductor nanowire FETs have been calibrated with available experimental results in our previous works [37, 38]. It is apparent from the plots of Fig. A2.1 that, the device current at room temperature would significantly depend on the net phonon scattering, if the (virtual) energy gaps between ground state and 1$^{st}$ excited state were below or around the thermal energy, *i.e.*, from < 25 meV to ~200 meV. However, the effect of such phonon scattering becomes negligible for an energy gap > 200 meV, and consequently, the device current becomes equal to that obtained without considering any phonon scattering in the present device (as can be seen in Fig. 4). Therefore, the effect of phonon scattering even at room temperature in the highly quantized VTQDs within GaAs nanowire device considered in the present work is negligibly small and can be neglected.

### Appendix-III: Current in time domain and dephasing time

It is interesting to note that an electron exhibits coherent oscillation in a double quantum dot device due to the inter-dot coupling, which has a natural decay time due to energy-time uncertainty of such coupling potential that depends on the material properties and device dimensions. Therefore, even for ballistic transport, in the absence of phonon scattering or for its insignificant influence (*i.e.*, the present case as shown in the Appendix-II), the qubit superposed state suffers dephasing after 'initialization', once the pulse at the relevant gate (*i.e.*, gate-2) is applied for 'manipulation'. Thus, while 'measurement' is performed by applying a small drain bias, the amplitude of tunneling current decays in time as shown in Fig. 4. Such time dependence can be obtained as follows [37]:

The Fourier transform of Eq.(3) into time domain gives rise to,

$$\frac{1}{2\pi\hbar}\int_{-\infty}^{\infty}(E - H_{Iso} - \Sigma_S(E) - \Sigma_D(E))[G(E)]e^{-\frac{i}{\hbar}Et}dE = \delta(t) \quad \text{(A3-1)}$$

which, considering the dimensionless retarded Green's function to be $G^{Rt}(t)$, leads to,

$$\left(i\hbar\frac{d}{dt} - H_{Iso}\right)G^{Rt}(t) = \int(1 + (\Sigma_S(E) + \Sigma_D(E))G(E))e^{-\frac{i}{\hbar}Et}dE \quad \text{(A3-2)}$$

$$\Rightarrow G^{Rt}(t) = \left(i\hbar\frac{d}{dt} - H_{Iso}\right)^{-1}\int(1 + (\Sigma_S(E) + \Sigma_D(E))G(E))e^{-\frac{i}{\hbar}Et}dE$$

$$\Rightarrow G^{Rt}(t) = \int G_{Iso}(E)(1 + (\Sigma_S(E) + \Sigma_D(E))G(E))e^{-\frac{i}{\hbar}Et}dE$$

$$\Rightarrow G^{Rt}(t) = F.T._{E\to t}[G_{Iso}(E)] + F.T._{E\to t}[G_{Iso}(E)\Sigma_S(E)G(E)] + + F.T._{E\to t}[G_{Iso}(E)\Sigma_D(E)G(E)]$$

(A3-3)

where, $G_{Iso}(E)$ is the Green's function for isolated nanowire without considering its coupling with the reservoirs (*i.e.*, source/drain). It is interesting to note that the first term in the expression (A3-3) of retarded Green's function represents the contribution of isolated nanowire whereas the second and third terms are associated with its coupling with the source and drain, respectively. Therefore, the time dependence of drain current during 'measurement' is obtained from the current amplitude multiplied by the retarded Green's function for drain as,

$$I = I_0(F.T._{E\to t}[G_{Iso}(E)\Sigma_D(E)G(E)]) \quad \text{(A3-4)}$$

Such retarded Green's function for drain having both real and imaginary parts leads to an oscillatory decay of the current pulse as plotted in Fig. 4. The pulse repetition time in the present dual-gate nanowire FET device is obtained by calculating the lifetime of electron pulse entering QD-1, by taking an average of time over the norm of retarded Green's function [37] for source (*i.e.*, 2[nd] term of Eq. A3-3); whereas the dephasing time is calculated by taking the average of time over the norm of retarded Green's function for drain (*i.e.*, 3[rd] term of Eq. A3-3) up to the limit of pulse repetition time.

## Data availability statement

The datasets generated and analysed during the current study are available from the corresponding author on reasonable request.

## Conflict of interest

The authors have no conflicts to disclose.